\documentclass[aip,jcp,twocolumn,superscriptaddress]{revtex4}%
\usepackage{epsfig,amssymb,amsmath,amsthm,amsfonts,amsbsy,mathrsfs}
\usepackage[linesnumbered,ruled]{algorithm2e}
\usepackage{multirow}

\newcommand{\br}{\mathbf{r}}
\newcommand{\bR}{\mathbf{R}}
\newcommand{\bk}{\mathbf{k}}

\newcommand{\vt}{\vec{\theta}}

\begin{document}

\title{Simulating periodic systems on quantum computer}

\author{Jie Liu}
%\thanks{liujie86@ustc.edu.cn (Jie Liu)}
\affiliation{Hefei National Laboratory for Physical Sciences at the
Microscale, University of Science and Technology of China, Hefei, Anhui 230026,
China}

\author{Lingyun Wan}
\affiliation{Hefei National Laboratory for Physical Sciences at the
Microscale, University of Science and Technology of China, Hefei, Anhui 230026,
China}

\author{Zhenyu Li}
\thanks{zyli@ustc.edu.cn (Zhenyu Li)}
\affiliation{Hefei National Laboratory for Physical Sciences at
the Microscale, University of Science and Technology of China,
Hefei, Anhui 230026, China}

\author{Jinlong Yang}
\thanks{jlyang@ustc.edu.cn (Jinlong Yang)}
\affiliation{Hefei National Laboratory for Physical Sciences at the
Microscale, Department of Chemical Physics, and Synergetic
Innovation Center of Quantum Information and Quantum Physics,
University of Science and Technology of China, Hefei, Anhui 230026,
China}

\date{\today}

\begin{abstract}
%Recently, several efficient quantum algorithms have been proposed for quantum chemistry
%simulation on near-term noisy intermediate-scale quantum devices. While most existing
%algorithms, such as quantum phase estimation (QPE) and variational quantum eigensolver (VQE),
%focus on molecular systems or model systems. As rapid progresses in quantum devices, it
%is necessary to explore more potential applications of quantum computing.
The variational quantum eigensolver (VQE) is one of the most appealing quantum algorithms to
simulate electronic structure properties of molecules on near-term noisy intermediate-scale 
quantum devices. In this work, we generalize the VQE algorithm for simulating periodic systems. 
However, the numerical study of an one-dimensional (1D) infinite hydrogen chain using 
existing VQE algorithms shows a remarkable deviation of the ground state energy with respect 
to the exact full configuration interaction (FCI) result. Here, we present two schemes
to improve the accuracy of quantum simulations for periodic systems. The first one is a modified VQE algorithm,
which introduces an unitary transformation of Hartree-Fock orbitals to avoid the complex Hamiltonian.  
The second one is a {\it Post}-VQE approach combining VQE with the quantum subspace expansion approach (VQE/QSE). 
Numerical benchmark calculations demonstrate that both of two schemes provide an accurate 
description of the potential energy curve of the 1D hydrogen chain. In addition, excited states computed with 
the VQE/QSE approach also agree very well with FCI results.
\end{abstract}

\maketitle

\section{Introduction}
A well-defined, accurate and efficient electronic structure method
is critical for interpreting material properties and for the prediction and
designing of novel materials. Over years, a huge amount of effort has been devoted to 
systematically improve the computational accuracy for material simulations.
%many approaches have been proposed to approximately solving the many-electron Schr\"odinger equation.
%A well-defined, accurate and efficient electronic structure method  for material simulations is
%one of the major open challenging problems in quantum physical and chemistry community.
Density functional theory (DFT) is a very powerful and elegant first principles method
to explore ground-state properties of solids in materials science and condensed matter physics while
the accuracy of DFT calculations strongly depends on exchange-correlation 
functional approximations~\cite{CohMorYan08,ZhaTru08a,CohMorYan12}.
Recently, wave function based periodic electronic structure methods, such as second-order M\o{}ller-Plesset perturbation theory
and coupled-cluster (CC) theory with single and double excitations (CCSD), have been 
successfully applied to problems in solids and low-dimensional nanomaterials
~\cite{SunBar96,AyaKudScu01,ShuDolFul99,GilAlfGir08,NolGilAlf09,BooGruKre13,McCSunCha17,HumTsaGru17}. 
%a systematic improvement toward the exact full configuration interaction (FCI) method 
%For example,coupled-cluster theory with single and double excitations (CCSD) plus perturbative triples
%correction, which scales like the seventh power of the system size,
%is known as the gold standard in quantum chemistry~\cite{RilPitJur10}. 
Although wave function based methods offer a systematic approach for 
solving the many electron Schr\"odinger equation, their computational cost 
for accurately capturing strong electronic correlation effects 
%including magnetic, structural, conductive and superconductive phase transitions 
is often prohibitive. For example, the exact full configuration interaction (FCI) method scales exponentially.
%Therefore, accurately and efficiently solving the many electron Schr\"odinger equation, 
%is still an open problem in computational and material science.

The recent advent of quantum computing provides a new pathway for solving many electron Schr\"odinger equation in polynomial 
time~\cite{BraKit02,GeoAshNor14,Pre18,CaoRomOls19,McAEndAsp20}. With rapid progresses in quantum chemistry based quantum algorithms, 
experimental studies of
molecular ground-state and excited-state properties have been extensively
performed~\cite{DuXuPen10,PerMcCSha14,WecHasTro15,MalBabKiv16,HarMon17,ReiWieSvo17,KanMezTem17,HemMaiRom18,SanWanGen18,GriEcoBar19}.
For example, Du {\it et al.} reported a quantum phase estimation (QPE) simulation of the ground-state energy of the hydrogen molecule
using NMR~\cite{DuXuPen10}.
Peruzzo {\it et al.} firstly proposed and realised the variational quantum eigensolver 
(VQE) algorithm on a photonic quantum processor
for computing the ground-state energy of HeH$^+$~\cite{PerMcCSha14}. 
Colless {\it et al.} used a superconducting-qubit-based
processor to apply the quantum subspace expansion (QSE) approach to the H$_2$ molecule,
extracting both ground and excited states without
the need for additional minimization~\cite{ColRamDah18}. 

The QPE ~\cite{AspDutLov05} and VQE algorithms~\cite{PerMcCSha14}
are two leading quantum algorithms for solving electronic structure problems on a
quantum computer. The standard QPE algorithm evolves in time a quantum state under the 
Hamiltonian $\hat{H}$ of interest, which offers an exponential speedup for determining
the molecular spectra over classical methods. However, The practical implementation of QPE requires
large, error-corrected quantum computer, which is believed to be out of
reach in near-term quantum devices. On the other side, VQE provides an alternative quantum algorithm for
near-term noisy intermediate-scale hardware because it requires a much
shorter coherent time and can be implemented with massive parallelization~\cite{MalBabKiv16}.

VQE applies the Rayleigh-Ritz variational principle to optimize the parameterized wave function, 
which finally minimizes the total energy functional.
The variational optimization procedure is a hybrid quantum-classical algorithm, that is, the evaluation of 
various physical properties in terms of the expectation value of operators, 
such as the energy and gradient, is performed on the quantum computer
and the update of parameters is performed on the classical
computer~\cite{WhiBiaAsp11,McCRomBab16,McCKimCar17,RomBabMcC18,ColRamDah18}.
Given the wave function, the expectation value of operators is obtained by repeating the measurement 
many times and taking an average of measurement values.
Integrated with the quantum computer, the state preparation and measurement of Pauli operators 
can be carried out in polynomial time.    
Therefore, it is possible to efficiently compute the energy or gradient on 
the quantum computer even when the wave function involves exponential configurations.
Considering the superiority of VQE on near-term quantum devices, it has been widely used to
solve various electronic structure problems~\cite{CaoRomOls19}.
%Despite of the huge success of VQE on simulating molecular properties, 
However, the application of VQE to periodic systems is still lacking.

In this work, we generalize the VQE algorithm to periodic systems.
The wave function ansatzes used in this work include the unitary coupled cluster (UCC)
ansztz~\cite{Eva11,PerMcCSha14} and the 
Adaptive Derivative-Assembled Pseudo-Trotter ansatz (ADAPT)
recently proposed by Grimsley {\it et al.}~\cite{GriEcoBar19}.
%ADAPT-VQE requires a small number of parameters, that is shallow circuit depth,
%to reach the same accuracy as Unitary CCSD. 
With the periodic boundary condition, Hartree-Fock (HF) orbitals at sampling $k$-points are defined 
in the complex number space in order to satisfy the translational symmetry.
Given anti-Hermitian excitation operators used in UCC and ADAPT-VQE, coefficients of operators are assumed to be 
real in order to generate an unitary transformation. 
Therefore, the complex wave function variationally optimized in the real parameter space
converges to a local minimum as discussed later in this work.
%The total energy and parametrized wave function are optimized following the variational principle. 
%However, as the $k$-point sampled Bloch molecular orbitals are intro
%As the $k$-point sampled Hamiltonian is introduced in calculations of periodic systems,
%the anti-Hermitian contracted Schr\"odinger equation (ACSE)~\cite{Maz06,Maz07} does not necessarily vanishes as 
%the total energy functional is minimized. 
%The residual error of ACSE results in a remarkable deviation of the ground-state energy. 
In order to overcome this problem, we propose a modified VQE algorithm, named the VQE-K2G approach,
which converts HF orbitals at sampling $k$-points in an 
unit cell into real orbitals at $\Gamma$-point in the corresponding supercell. 
After that, the wave function and Hamiltonian are also defined in the real space.
Therefore, the VQE-K2G algorithm for periodic systems is expected to be as
accurate as VQE for molecular systems.
%In the K2Gamma scheme,
%the imaginary part of ACSE vanishes throughout the variational procedure and 
%the residual error of the anti-Hermitian part of CSE is minimized in the variation optimization procedure.
In addition, it is possible to improve the accuracy of VQE
by combining it with QSE (VQE/QSE)~\cite{SanWanGen18}, in 
which a reference state is prepared with VQE and the ground state wave function 
is obtained by diagonalizing the Hamiltonian sampled on the linear-response space of the 
reference state. VQE/QSE is expected to offer a good estimation of the exact wave function if 
VQE can prepare a reasonable reference state.

This paper is organized as follow. Section~\ref{sec:theory} gives a
brief description of the theoretical methodology,
covering periodic Hartree-Fock method, VQE algorithms with UCC and ADAPT ansatzes, 
the VQE algorithm for periodic systems and the VQE/QSE approach.
In section~\ref{sec:results}, we first compute the ground-state potential energy
curve of an equispaced one-dimensional (1D) infinite hydrogen chain and analyze errors 
for different wave function ansatzes. We then assess the accuracy of two schemes, 
VQE-K2G and VQE/QSE, for ground-state calculations. Finally, we compute the potential energy curve 
of the first excited state for the 1D hydrogen chain with the VQE/QSE approach. 
A summary and outlook is given in Section~\ref{sec:conclusion}.

\section{Theory}\label{sec:theory}

\subsection{Periodic Hartree-Fock method}
For the periodic system using atom-centered basis sets, Bloch atomic orbitals (BAO) are defined as
\begin{equation}
    \chi_{\mu \bk}(\br) = \frac{1}{\sqrt{N}}\sum_{\bR_n} e^{i\bk \cdot \bR_n} \chi_\mu (\br-\bR_n)
\end{equation}
where $\bR_n$ is the translation vector and $\bk$ is a crystal momentum vector sampled in the unit
cell. $N$ is the number of unit cells. 
%when $N \rightarrow \infty$, $ \chi_{\mu k}(\br)$ is fully periodic
%with respect to all lattice translations and therefore satisfies the Bloch theorem.
HF orbitals at $\bk$ are expanded as a linear combination of BAOs,
\begin{equation}\label{eq:k-wf}
    \phi_{p\bk}(\br) = \sum_{\mu} C_{\mu p}(\bk) \chi_{\mu\bk}(\br)
\end{equation}
which is also known as the Hartree-Fock-Roothaan approximation in first principles molecular calculations.
Given HF orbitals, the corresponding one- and two-electron integrals can be computed~\cite{VanKraMoh05}.
%there are several strategies to compute the corresponding one- and two-electron integrals. Here, 
%we compute kinetic, ionic potential and Coulomb 
%integrals with the mixed Gaussian and plane wave strategy~\cite{VanKraMoh05}. 
The core electron potential is
represented with the norm-conserving HGH pseudopotential~\cite{KleByl82,GoeTetHut96,HarGoeHut98}, 
which removes the Coulomb singularity at the origin.
%The two-electron integrals are computed through a fast Fourier transform with auxiliary 
%plane-wave basis set.

For each $\bk$, the Hartree-Fock eigenvalue equation in the representation of BAOs
is expressed as,
\begin{equation}
    F(\bk)C(\bk) = S(\bk)C(\bk)E(\bk)
\end{equation}
The Fock and overlap matrix elements are given by
\begin{equation}
   \begin{split}
    F_{\mu\nu}(\bk) &= T_{\mu\nu}(\bk) + V^{\mathrm{PP}}_{\mu\nu}(\bk) + J_{\mu\nu}(\bk) + K_{\mu\nu}(\bk) \\
    S_{\mu\nu}(\bk) &= \int_\Omega \chi_{\mu\bk}^*(\br) \chi_{\nu\bk}(\br) d \br
    \end{split}
\end{equation}
Here $\mathbf{T}$ is the kinetic energy, $\mathbf{V}^{\mathrm{PP}}$ is the pseudopotential, $\mathbf{J}$ is the Coulomb integral and $\mathbf{K}$ is the
exchange integral. In the real space, these matrices are given by
\begin{equation}
    \begin{split}
        &T_{\mu\nu}(\bk) = -\frac{1}{2}\int_\Omega \chi_{\mu\bk}^*(\br) \bigtriangledown^2_{\br} \chi_{\nu\bk}(\br) d \br \\
        &J_{\mu\nu}(\bk) = \int \int_\Omega \chi_{\mu\bk}(\br) \frac{\rho(\br',\br')}{|\br-\br'|} \chi^*_{\nu\bk}(\br) d \br d \br' \\
        &K_{\mu\nu}(\bk) = \int \int_\Omega \chi_{\mu\bk}(\br) \frac{\rho(\br,\br')}{|\br-\br'|} \chi^*_{\nu\bk}(\br')  d \br d \br'
    \end{split}
\end{equation}
where $\Omega$ indicates that the real space integration is performed in the unit cell.
The density matrix can be obtained by averaging over sampling $k$-points in an unit cell
\begin{equation}
    \rho(\br,\br') = \frac{1}{N_\bk} \sum_\bk \sum_i^{N_{o}} \phi_{i\bk}(\br) \phi^*_{i\bk} (\br')
\end{equation}
where $N_k$ is the number of $k$-points and $N_o$ is the number 
of occupied electrons.

\subsection{VQE algorithm}

In the VQE algorithm, one key ingredient is the wave function ansatz, which prepares an electronic state 
with a few parametrized unitary operators~\cite{McCRomBab16},
\begin{equation}\label{eq:unitary_transformation}
    |\psi(\vec{\theta})\rangle = U(\vec{\theta})|\psi_0\rangle,
\end{equation}
where $|\psi_0\rangle$ is the reference state. The parametrized wave function is
optimized through Rayleigh-Ritz variational principle
\begin{equation}\label{eq:RR}
    E_0 = \min_{\vt} \{ \langle \psi(\vec{\theta}) | \hat{H} | \psi(\vec{\theta}) \rangle \}.
\end{equation}
The Hamiltonian in the second-quantized representation is expressed as
\begin{equation}
\hat{H} = \sum_{pq} h^p_q \hat{T}^p_q + \frac{1}{2} \sum_{pqrs} h^{pq}_{rs} \hat{T}^{pq}_{rs}
\end{equation}
where $h^p_{q}$ is the one-electron integral, including kinetic energy and ionic potential
(pseudopotential in this work) and $h^{pq}_{rs}$ is two-electron integral
\begin{equation}
    h^{pq}_{rs} = \int \int \phi_p^*(\br_1)\phi_q^*(\br_2)\frac{1}{|\br_1-\br_2|}\phi_r(\br_2) \phi_s(\br_1)d \br_1 d\br_2.
\end{equation}
The general one- and two-body excitation operators are defined as~\cite{Noo00,MukKut04}
\begin{equation}
    \begin{split}
        \hat{T}^p_q &= \hat{a}_p^\dag \hat{a}_q \\
        \hat{T}^{pq}_{rs} &= \hat{a}_p^\dag \hat{a}_q^\dag \hat{a}_r \hat{a}_s
    \end{split}
\end{equation}
$a_p^\dag$ and $a_p$ are the second-quantized creation and annihilation operators,
satisfying the anticommutation relation.

\subsubsection{Unitary coupled cluster ansatz}
Unitary coupled cluster (UCC) ansatz is a common component in quantum variational algorithms~\cite{Kut82,BarKucNog89,TauBar06,Eva11,HarShiScu18}. Different
from the traditional CC (tCC) theory, the UCC energy and wave function are variationally 
determined according to Eq.~\eqref{eq:RR}.
UCC wave function is defined as
\begin{equation}\label{eq:ucc}
    |\psi\rangle = e^{T-T^\dag} |\psi_0 \rangle.
\end{equation}
A cluster operator for unitary CCSD (UCCSD) is expressed as~\cite{TauBar06}
\begin{equation}
        \hat{T} = \sum_{ai} t^a_i \hat{T}^a_i + \frac{1}{4}\sum_{abij} t^{ab}_{ij} \hat{T}^{ab}_{ij}.
\end{equation}
Recently,
a generalized UCCSD (UCCGSD) wave function has been introduced in the VQE algorithm~\cite{Ron03,Maz04,Nak04,MukKut04,PerMcCSha14}
\begin{equation}
    \hat{T} = \frac{1}{2}\sum_{pq} t^p_q \hat{T}^p_q + \frac{1}{4}\sum_{pqrs} t^{pq}_{rs} \hat{T}^{pq}_{rs}.
\end{equation}
Here $a,b,\ldots$ indicate virtual orbitals; $i,j,\ldots$ indicate occupied orbitals; 
and $p,q,\ldots$ indicate general orbitals. 

Although UCC is more robust than tCC,
neither UCCSD or UCCGSD can be expanded using the Baker-Campbell-Hausdorff
(BCH) formula at finite order. Therefore, an exact implementation of UCC on
a classical computer scales exponentially.
While the UCC wave function can be easily prepared on a quantum computer even if the reference state is a
multiconfigurational state. Recently, the UCC ansatz has been widely 
used in experimental and theoretical 
chemistry simulations with the VQE algorithm~\cite{PerMcCSha14,SheZhaZha17}.

At convergence, the stationary of the energy with respect to 
parameters is expressed as
\begin{equation}\label{eq:uccsd_grad}
       \frac{\partial  \langle \psi(\vec{t}) | \hat{H} | \psi(\vec{t}) \rangle }{\partial t_u} = 0,
\end{equation}
where $t_u$ is the coefficient of anti-Hermitian operators 
$\tau_u \in \{\hat{T}^p_q-\hat{T}^q_p, \hat{T}^{pq}_{rs}-\hat{T}^{sr}_{qp}\}$ for UCCGSD
and $\tau_u \in \{\hat{T}^a_i-\hat{T}^i_a, \hat{T}^{ab}_{ij}-\hat{T}^{ji}_{ba}\}$ for UCCSD.
Recent numerical studies of small molecules with minimum basis sets
demonstrate UCCGSD is far more robust and accurate than UCCSD~\cite{LeeHugHea19}. The difference between UCCSD and UCCGSD 
is even more significant for periodic numerical simulations as shown later in this work.
 
On quantum computers, the implementation of UCC requires
a decomposition of the exponentiated cluster operator into one- and two-qubit gates using an approximate scheme, such as the
Trotter-Suzuki decomposition~\cite{PouHasWec15,BabMcCWec15,GriClaEco20},
\begin{equation}
    e^{\hat{A}+\hat{B}} \approx \left( e^{\hat{A}/k} e^{\hat{B}/k} \right)^k.
\end{equation}
The UCC wave function with Trotterization is defined as
\begin{equation}\label{eq:ucc_trotter}
    |\psi\rangle = \prod_k^\infty \prod_u^{N_u} e^{\frac{t_u}{k}\tau_u} |\psi_0 \rangle.
\end{equation}
Therefore, the accuracy of UCC-VQE simulations strongly depends on the Trotter formula used,
the number of Trotter steps, and the time-ordered sequence of operators in the
UCC ansatz. In principle, the lower-order Trotter-Suzuki decomposition will result in larger
error. However, given the wave function expression in Eq.~\eqref{eq:ucc_trotter} , 
the optimization of the wave function in the parameter space is able to 
cancel part of error and give a promising estimation of the ground-state energy.

%In the early work, a single Trotter step was shown to yield accurate results.
%Therefore, the UCCSD operator must be broken up into a time-ordered sequence of few (one or
%two) particle operators. Because the generalized single and double excitation operators
%do not commute, the use of a truncated Trotter expansion
%represents an approximation to the underlying UCCSD ansatz,
%and recent work has shown clearly that this will indeed
%affect the results. As a result, determining a suitable (or even optimal)
%ordering of the operators in the UCCSD Ansatz may be an
%interesting and important area of future research.

\subsubsection{ADAPT ansatz}
Grimsley {\it et al.} recently proposed to approximate the exact wave function
as an arbitrarily long product of general one- and two-body exponentiated operators~\cite{GriEcoBar19},
\begin{equation}\label{eq:wave0}
    | \psi (\vec{\theta}) \rangle = \prod_k^{N_k} \prod_u^{N_u} e^{\theta_u(k) \tau_u} |\phi_0 \rangle.
\end{equation}
%Eq~\eqref{eq:wave0} is very similar with Eq~\eqref{eq:ucc_trotter} expect that 
%parameters of ADAPT-VQE wave function in each replica $k$ are assumed to be different. 
%Therefore, the numerical results from UCC and ADAPT-VQE may be close to each other as shown 
%in Section~\ref{results} but they are two different approaches.
%Algorithm~\ref{alg:vqe} shows the variational optimization procedure for minimizing the 
%total energy.
In order to generate a maximally compact sequence of operators at
convergence, the operator, $\tau(k)$, with the largest absolute {\it pre-estimated} gradient 
instead of all operators in the operator pool $\mathcal{O}$
is used to update the wave function ansatz in the $k$-th iteration.
This indicates $N_u=1$ in Eq~\eqref{eq:wave0}. The wave function is iteratively updated with
\begin{equation}\label{eq:wave}
    | \psi (k) \rangle =  e^{\theta(k) \tau(k)} |\psi(k-1) \rangle
\end{equation}
where $|\psi(0)\rangle = |\phi_0\rangle$ is the reference state.
The energy functional in the $k$-th iteration is minimized by
\begin{equation}\label{eq:energy}
    E(k) = \min_{\{\theta(l)\}_{l=1}^{k}} \{ \langle \psi(k) | \hat{H} | \psi(k) \rangle \}.
\end{equation}
The gradient of the energy functional with respect to parameters $\{\theta(l)\}_{l=1}^{k}$ is 
formulated as
\begin{equation}\label{eq:acse_grad}
\begin{split}
    G_{l} &= \frac{\partial E(k)}{\partial \theta(l)} \\
    &= 2\Re\left(\langle \psi(k) | \hat{H} \prod_{m=l+1}^{k} \left(e^{\theta(m) \tau(m)}\right) \tau(l) \right. \\
    & \quad \quad \quad \quad\left.\prod_{n=1}^l \left(e^{\theta(n) \tau(n)}\right) |\phi_0 \rangle \right).
    \end{split}
\end{equation}
The convergence criteria is defined as 
\begin{equation}\label{eq:acse_conv}
    |\vec{R}|_2 = \sqrt{\sum_u |R_u|^2} < \epsilon
\end{equation}
where $R_u$ is the {\it pre-estimated} gradient for the next iteration.
For example, in the $(k+1)$-th iteration, $R_u$ is defined with the wave function optimized in the $k$-th iteration,
\begin{equation}\label{eq:acse_res}
\begin{split}
    R_u &= G_{k+1}|_{\theta_{k+1}=0,\tau(k+1)=\tau_u} \\
    &= \langle \psi(k)| [\hat{H},\tau_u] |\psi(k) \rangle.
    \end{split}
\end{equation}
The variational optimization procedure for the ADAPT-VQE algorithm is summarized in Algorithm~\ref{alg:vqe}.
%At convergence, ADAPT-VQE wave function should satisfy
%\begin{equation}\label{eq:acse_conv}
%    \sum_u |\langle \psi | [\hat{H},\tau_u] |\psi\rangle|^2 < \epsilon^2
%\end{equation}
%In this work, ADAPT(m) indicates an ADAPT-VQE calculation with
%$\epsilon=10^{-k}$. If a small enough convergence thresh $\epsilon$
%is set, Eq.\eqref{eq:acse_conv} is equivalent to Brillouin
%condition~\cite{Maz07}.
%which means ADAPT-VQE converges with the same condition as
%UCCGSD. We can expect the same result from UCCGSD and ADAPT-VQE with a tight convergence criterion.

\begin{algorithm}[!htb]
\label{alg:vqe}
\SetAlgoNoLine
\caption{The ADAPT-VQE algorithm for optimizing the wave function and the energy.}

\KwIn{Reference state $|\psi_0\rangle$ and Hamiltonian $\hat{H}$;}

\KwOut{The energy and wave function of the target state;}

Prepare the initial wave function $|\Psi \rangle = |\psi_0 \rangle$ in qubit representation\;

Define the operator pool $\mathcal{O}$\;

Initialize the operator set $\vec{\tau}=\{\}$ and parameters $\vec{\theta}=\{0\}$\;

\While {$|\vec{R}|_2 > \epsilon$}
{
Compute \{$\vec{R}$\} with Eq.~\ref{eq:acse_res} for all $\tau_u \in \mathcal{O}$\;

$\vec{\tau} \gets \{\vec{\tau}, \tau_u\} $ with $|R_u|$ being the largest absolute {\it pre-estimated} gradient and $\vec{\theta}=\{\vec{\theta},0\}$\;

Define the new wave function with Eq.\eqref{eq:wave} and the new energy functional with Eq.\eqref{eq:energy}\;

Optimize parameters $\vec{\theta}$\;
}

Return E($\vec{\theta}$) and $|\psi(\vec{\theta})\rangle$.

\end{algorithm}

\subsection{VQE algorithm for periodic systems}\label{sec:periodic}
To extend VQE algorithms for periodic systems, we define the general one- and two-body
operator pool with Bloch orbitals~\cite{PieKowFan03,Dav03,Nak04},
\begin{equation}\label{eq:pbc_t}
    \begin{split}
        \hat{T}^{\tilde{p}}_{\tilde{r}} &= \hat{a}_{\tilde{p} }^\dag \hat{a}_{\tilde{r}} \\
        \hat{T}^{\tilde{p}\tilde{q}}_{\tilde{r}\tilde{s}} &= \hat{a}_{\tilde{p} }^\dag \hat{a}_{\tilde{q} }^\dag \hat{a}_{\tilde{r}}\hat{a}_{\tilde{s}}
    \end{split}
\end{equation}
where $\tilde{p}=p\bk_p$. 
The Hamiltonian is summed over sampling $k$-points in an unit cell,
\begin{equation}\label{eq:bloch_ham}
\hat{H} = \sum_{\tilde{p}\tilde{r}}^\prime h^{\tilde{p}}_{\tilde{r}} \hat{T}^{\tilde{p}}_{\tilde{r}} + \frac{1}{2} \sum_{\tilde{p}\tilde{q}\tilde{r}\tilde{s}}^\prime h^{\tilde{p}\tilde{q}}_{\tilde{r}\tilde{s}} \hat{T}^{\tilde{p}\tilde{q}}_{\tilde{r}\tilde{s}}.
\end{equation}
Because the Hamiltonian satisfies the translational symmetry based on the periodic boundary condition, general one- and two-body excitation
operators must conserve crystal momentum, 
\begin{equation}\label{eq:double}
\sum_p \bk_p - \sum_r \bk_r = \mathbf{G}_m
\end{equation}
where $\bk_p$ and $\bk_r$ are crystal momenta of creation operators and annihilation operators, respectively. $\mathbf{G}_m$ is a reciprocal lattice vector. 
The primed summation in Eq.~\eqref{eq:bloch_ham} indicates that one of the orbital momenta is fixed according to Eq.~\eqref{eq:double}. 

In order to analyze the difference between real and complex HF orbitals, we firstly introduce
the contracted Schr\"odinger equation (CSE), which can be derived by contraction of the Schr\"odinger equation onto the space of
two particles
\begin{equation}
    \langle \psi | \hat{T}_u \hat{H} | \psi \rangle = E \langle \psi | \hat{T}_u | \psi \rangle.
\end{equation}
The anti-Hermitian CSE (ACSE) is expressed as
\begin{equation}\label{eq:acse_general}
    \langle \psi | [ \hat{T}_u, \hat{H} ] | \psi \rangle = 0,
\end{equation}
which is also known as the Brillouin condition~\cite{Maz06,Maz07}. 
The ACSE can be written as a sum of the real part (ACSE-Re)
\begin{equation}\label{eq:acse_real}
    \langle \psi | [ \tau_u, \hat{H} ] | \psi \rangle = 0,
\end{equation}
and the imaginary part (ACSE-Im)
\begin{equation}\label{eq:acse_imag}
    \langle \psi | \{ \tau_u, \hat{H} \} | \psi \rangle = 0.
\end{equation}
Therefore, for the real wave function, the variationally optimized  
ADAPT-VQE wave function is exactly the solution of ACSE. This agrees with 
the conclusion that ACSE enforce stationary of the energy with respect to 
a sequence of unitary transformation of the reference state. 
%Given the convergence 
%condition of ADAPT-VQE as indicated in Eq.~\eqref{eq:acse_conv}, the ADAPT-VQE wave function 
%satisfies the ACSE at convergence if a small enough $epsilon$ is set for optimization.
%As the Bloch wave function is introduced, the variational principle only guarantees
%that the real part of ACSE vanishes at convergence. 
%The residual error of ACSE, which uniquely determines the deviation of
%the total energy if ACSE converges below a small thresh, will strongly depend on the optimized wave function.  
As the Bloch wave function is introduced, Eq.~\ref{eq:acse_real} 
is still satisfied according to the convergence criteria in Eq.~\ref{eq:acse_conv} while 
the residual error of ACSE-Im that results in the deviation of the energy is not minimized in the variational optimization procedure.
In order to remove the residual error of ACSE-Im, we transform
HF orbitals at sampling $k$-points in an unit cell into
orbitals at $\Gamma$-point in the corresponding supercell. This generates a set of 
real wave function and Hamiltonian for the VQE algorithm. 
A brief introduction of this scheme, named K2G, is described as following.
HF orbitals in Eq.~\ref{eq:k-wf} can be rewritten as
\begin{equation}
     \phi_{i\bk}(\br) = \sum_{\tilde{\mu}} \tilde{C}_{\tilde{\mu} i\bk } \chi_{\tilde{\mu}}(\br)
\end{equation}
where
\begin{equation}
    \tilde{C}_{\tilde{\mu} i\bk} = \frac{1}{\sqrt{N}} e^{i\bk \cdot \bR_n} C_{\mu i}(\bk)
\end{equation}

and $\tilde{\mu}$ is the $\mu$-th atomic orbital in $n$-th "replica".
The Fock matrix with the supercell atomic orbital basis is expressed as
\begin{equation}
    F_{\tilde{\mu} \tilde{\nu}} = \sum_{i\bk} \tilde{C}_{\tilde{\mu} i\bk} E_{i \bk} \tilde{C}_{\tilde{\nu} i\bk}^\dag
\end{equation}
Diagonalizing the Fock matrix, we obtain the real HF orbitals expanded with the supercell atomic 
basis functions.

\begin{figure}[!htb]
\begin{center}
\includegraphics[width=0.45\textwidth]{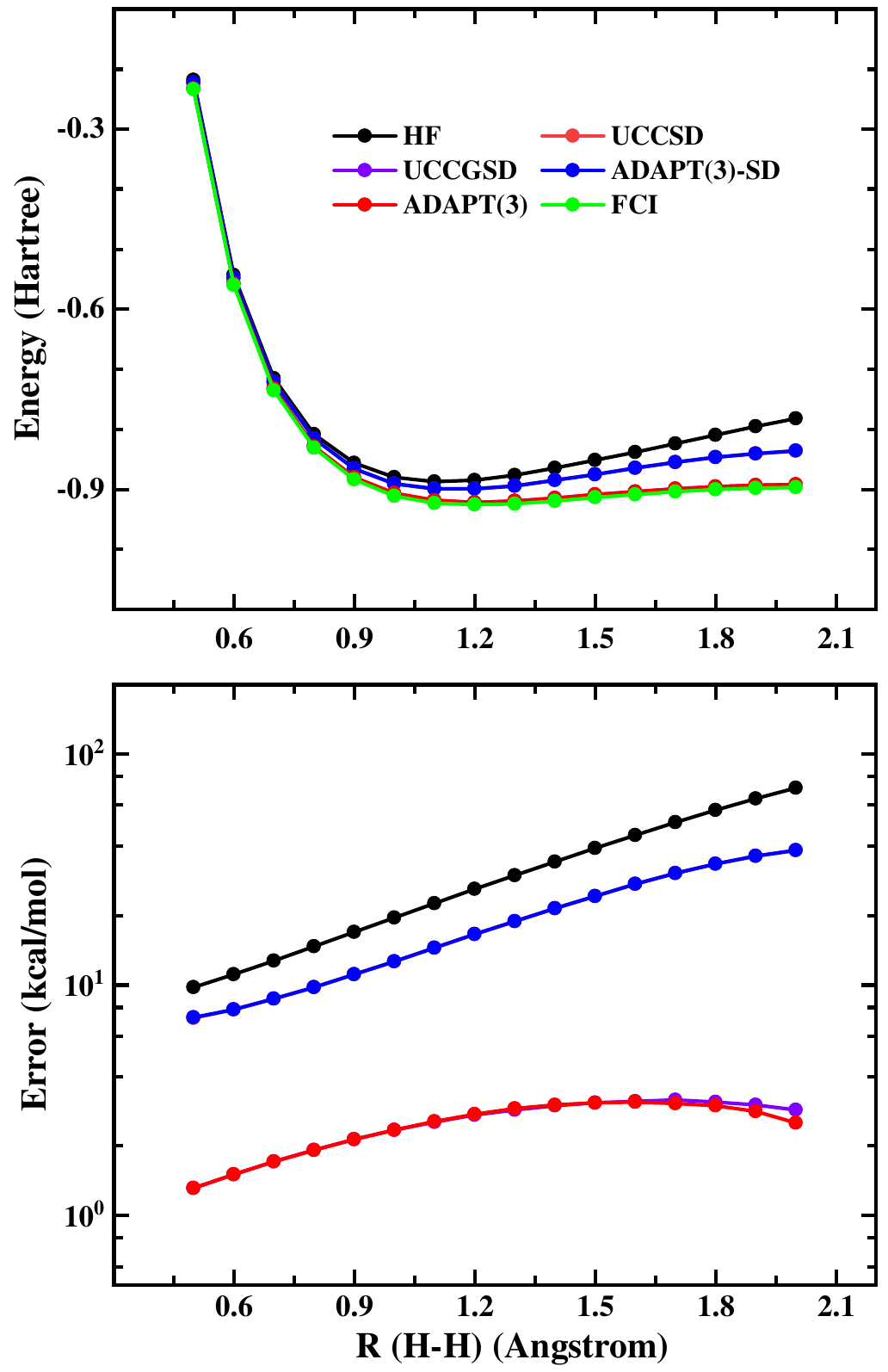}
\end{center}
\caption{The ground-state potential energy curve and absolute energy error
with respect to the FCI result for 1D hydrogen chain computed 
with UCC-VQE and ADAPT-VQE using HF orbitals at sampling k-points.} \label{fig:vqe_hx1}
\end{figure}

\begin{table*}[!htb] 
\centering \caption{Maximum absolute residual error (MARE) (in kcal/mol) of ACSE-Re and ACSE-Im for HF,
UCCSD-VQE, UCCGSD-VQE, ADAPT(3). mMARE indicates the mean MARE. }
%\begin{tabular}{|p{2cm}|p{2cm}|p{2cm}|p{2cm}|p{2cm}|p{2cm}|p{2cm}|p{2cm}|p{2cm}|}
\begin{tabular}{p{1.5cm}p{1.5cm}p{1.5cm}p{0.5cm}p{1.5cm}p{1.5cm}p{0.5cm}p{1.5cm}p{1.5cm}p{0.5cm}p{1.5cm}p{1.5cm}}
\hline \hline
\multirow{2}{*}{R} &\multicolumn{2}{c}{HF} && \multicolumn{2}{c}{UCCSD-VQE} &&  \multicolumn{2}{c}{UCCGSD-VQE} && \multicolumn{2}{c}{ADAPT(3)} \\
\cline{2-3}\cline{5-6}\cline{8-9}\cline{11-12}
& ACSE-Re & ACSE-Im &&  ACSE-Re & ACSE-Im && ACSE-Re & ACSE-Im && ACSE-Re & ACSE-Im \\
\hline
0.5 &19.23 &    4.82 &&18.69 &  5.03 && 0.02 &  8.28 && 0.02 &  8.21  \\
0.6 &16.64 &    4.66 &&16.01 &  5.08 && 0.03 &  7.11 && 0.02 &  7.12  \\
0.7 &14.86 &    4.78 &&14.10 &  5.46 && 0.03 &  6.39 && 0.02 &  6.38  \\
0.8 &13.61 &    4.89 &&12.71 &  5.67 && 0.02 &  5.89 && 0.03 &  5.89  \\
0.9 &12.70 &    4.99 &&11.66 &  5.86 && 0.02 &  5.58 && 0.06 &  5.60  \\
1.0 &12.04 &    5.07 &&10.92 &  6.04 && 0.03 &  5.35 && 0.04 &  5.38  \\
1.1 &11.56 &    5.14 &&10.97 &  6.25 && 0.03 &  5.15 && 0.09 &  5.21  \\
1.2 &11.21 &    5.20 &&11.06 &  6.45 && 0.03 &  4.96 && 0.05 &  4.99  \\
1.3 &10.96 &    5.25 &&11.16 &  6.65 && 0.03 &  4.76 && 0.06 &  4.79  \\
1.4 &10.80 &    5.28 &&11.23 &  6.85 && 0.05 &  4.62 && 0.05 &  4.63  \\
1.5 &10.67 &    5.32 &&11.22 &  7.05 && 0.05 &  4.44 && 0.05 &  4.44  \\
1.6 &10.65 &    5.33 &&11.07 &  7.25 && 0.09 &  4.30 && 0.07 &  4.30  \\
1.7 &10.81 &    5.35 &&10.72 &  7.45 && 0.14 &  4.10 && 0.05 &  4.14  \\
1.8 &11.03 &    5.35 &&10.08 &  7.62 && 0.18 &  4.02 && 0.06 &  4.00  \\
1.9 &11.24 &    5.34 &&9.14  &  7.73 && 0.25 &  3.85 && 0.08 &  3.83  \\
2.0 &11.47 &    5.33 &&7.93  &  8.35 && 0.13 &  3.50 && 0.04 &  3.82  \\
\hline
mMARE &12.47 &	5.13 &&11.79 &	6.55 &&	0.09 &	5.16 &&	0.05 &	5.17  \\
\hline \hline
\end{tabular}  \label{table1}
\end{table*}
\subsection{Quantum subspace expansion}\label{sec:qse}
The quantum subspace expansion approach has experimentally and theoretically proven to be one
of the most useful techniques on near-term noisy intermediate-scale quantum 
devices~\cite{McCKimCar17,ColRamDah18,SanWanGen18,PauAbhChu19}.
In the original implementation of QSE, the reference state is prepared with the 
UCCSD ansatz and excited states are obtained by diagonalizing the Hamiltonian sampled on the linear response space with single excitations~\cite{McCKimCar17}. It is straightforward to include higher-order 
excitation operators in QSE but at the expense of a sharp increase of measurements.

Given a set of linear-response excitation operators, $\{\hat{T}_u\}$, the configuration state space is defined as
\begin{equation}
    |\psi_u\rangle = \hat{T}_u |\psi\rangle.
\end{equation}
The QSE wave function can be expressed as a linear combination of configuration state functions
\begin{equation}\label{eq:qse_wf}
    |\psi^{\mathrm{QSE}}\rangle = \sum_u C_u |\psi_u\rangle.
\end{equation}
The energy and wave function of the ground and excited states can be obtained by solving a generalized
eigenvalue problem in the configuration state space
\begin{equation}\label{eq:qse}
    \mathbf{H}^{\mathrm{QSE}}\mathbf{C} = \mathbf{S}^{\mathrm{QSE}} \mathbf{C} \mathbf{E}
\end{equation}
with eigenvectors $\mathbf{C}$, and a diagonal matrix of eigenvalues $\mathbf{E}$. The Hamiltonian matrix elements
projected onto the configuration state space are given by
\begin{equation}
    H^{\mathrm{QSE}}_{u,v} = \langle \psi | \hat{T}_u \hat{H} \hat{T}_v | \psi \rangle.
\end{equation}
The overlap matrix elements are given by
\begin{equation}
    S^{\mathrm{QSE}}_{u,v} = \langle \psi | \hat{T}_u \hat{T}_v | \psi \rangle
\end{equation}
which is required because the configuration states are not necessarily orthogonal to each other.

%In the original formulation of VQE, the number of parameters will not exceed the number of 
%excitation operators ($N_{op}$) introduced in the wave function ansatz. Therefore,
%the optimized complex wave function can not satisfy the two set of converge condition with the 
%total dimension of 2$N_{op}$. In the QSE approach, the coefficient can be complex, 
%that is, the eigenvalue equation ~\ref{eq:qse_wf}
%is actually composed of two set of equations. When the VQE/QSE approach is accurate enough, 
%there is a big enough parameter space to optimize the wave function to satisfy
%Eq.~\eqref{eq:acse_general}.

The QSE approach is kind of inspired by the linear response method. 
In order to accurately describe the target state, either a well-defined reference state or the
inclusion of high-order excitations is necessary to obtain a converged result. The inclusion of high-order excitations 
indicates a polynomially increasing computational cost, which is prohibitive in medium- and large-size calculations.
The VQE algorithm, especially ADAPT-VQE, is able to efficiently generate a reasonable reference state
to approximate the target state even starting from a single-configurational Hartree-Fock state.
Here, we combine VQE with the QSE approach by truncating excitation operators up to second order 
and apply it to study the ground state and excited states.

\section{Results} \label{sec:results}
All calculations are performed with the modified ADAPT-VQE code~\cite{ADAPT-VQE}, which
uses OpenFermion~\cite{openfermion} for mapping fermion operators onto qubit operators
and PYSCF~\cite{pyscf} for all one- and two-electron integrals. The energy
and wave function are optimized with the
Broyden-Fletcher-Goldfarb-Shannon (BFGS) algorithm implemented in
SciPy~\cite{scipy}. Gradients are computed with
the finite difference approach for UCC-VQE and the analytical approach in Eq.~\eqref{eq:acse_grad} for
ADAPT-VQE. All UCC-VQE calculations are performed without Trotterization, 
that is, Eq.~\eqref{eq:ucc} is used in our classical numerical simulations.
All  full configuration interaction results used for
benchmark are obtained through explicitly diagonalizing the
Hamiltonian in {\it Hilbert} space of qubits. The operator pool is composed of 
the spin-adapted operators in order to avoid the spin contamination. 
SVZ basis set together with
GTH pseudopotential is used for all calculations.

Hereafter, for simplicity, we use ADAPT(m) to indicate an ADAPT-VQE calculation with
$\epsilon$=$10^{-m}$ defined in Eq.~\eqref{eq:acse_conv}. 
The prefix $\Gamma$ indicates a calculation with HF orbitals at $\Gamma$-point in the supercell.
For example, $\Gamma$-ADAPT(m) indicates an ADAPT(m) calculation using the K2G scheme.
ADAPT-SD and QSE-SD  
indicate that only one- and two-body excitations 
from occupied orbitals to virtual orbitals are involved in ADAPT and QSE, respectively.
%ADAPT(m)-SD is used for making comparison with UCCSD. 
As stated in Eq.~\eqref{eq:wave0}, we can specify $N_u > 1$ in ADAPT-VQE calculations. 
In this work, we update the wave function with 30 operators identified with the largest absolute {\it pre-estimated} gradients 
in each iteration. 

An 1D hydrogen chain with each hydrogen atom equispaced along a line is
an interesting model system to explore elemental physical phenomena in
modern condensed matter physics, such as an antiferromagnetic Mott
phase and an insulator-to-metal transition~\cite{MarClaFen19}. It also bridges the gap
between the simply Hubbard model and realistic bulk materials. 
Depending on different electronic spin alignments, the hydrogen chain can 
be paramagnetic, antiferromagnetic or ferromagnetic phase. 
In this work, we place two hydrogen atoms in an unit cell, which is 
the minimum model system used to study the magnetic properties.
Here, we benchmark the VQE algorithm for this 1D hydrogen chain model with one- and two-electron 
integrals obtained from a closed-shell 
Hartree-Fock calculation, which generates a paramagnetic state.
We vary the hydrogen-hydrogen bond length, $R$(H-H), 
and compute the potential energy curve of the
hydrogen chain with various wave function ansatzes.
In addition, we notice that a larger basis set and $k$-points 
sampling are required to accurately describe periodic system. 
However, the performance of various wave function ansatzes are assessed with respect 
to the FCI result computed with the same basis set and $k$-points. 
This will be not significantly affected by the accuracy of the theoretical method.
In the following, all calculations are performed with $1\times1\times4$ $k$-points, in which  
four $k$-points are sampled along the hydrogen chain and one $k$-point is
sampled along other two orthogonal directions. In the VQE-K2G approach, the corresponding supercell atomic basis functions are obtained by including eight hydrogen atoms in this supercell.

%\begin{figure}[!htb]
%\begin{center}
%\includegraphics[width=0.45\textwidth]{vqe-k-h1.pdf}
%\end{center}
%\caption{The deviations (in kcal/mol) of the total energy with respect to the FCI results by placing one hydrogen atom in an unit cell.} %\label{fig:vqe_hx2}
%\end{figure}

\subsection{Accuracy of VQE algorithms}\label{sec:conv}

In Figure~\ref{fig:vqe_hx1}, we study the ground-state potential energy curve and the absolute energy error with HF, 
UCCSD-VQE, UCCGSD-VQE, ADAPT(3)-SD, ADAPT(3) and FCI by varying $R$(H-H). HF fails to
reproduce the exact energy curve with the mean error (ME) of 32.37 kcal/mol because the
correlation effect totally misses in HF. UCCSD-VQE performs much better than HF with 
ME of 20.0 kcal/mol, but still significantly deviates from the exact
FCI result. UCCSD is able to treat most
weakly correlated systems while it suffers from the well-known problem of describing the
strong electronic correlation effect. For example, as $R$(H-H)
increases from 0.5 to 2.0 \r{A}, the absolute energy error of UCCSD-VQE increase 
from 7.19 to 38.16 kcal/mol. Note that the error of UCCSD-VQE in calculations of the 1D hydrogen chain is much 
larger than those in molecular calculations. For example,
the largest absolute energy error is only 9.19 kcal/mol in calculations of a challenging system N$_2$ 
with the STO-3G basis set~\cite{LeeHugHea19}. ADAPT-SD shares the same operator pool with UCCSD. 
Here, ADAPT(3)-SD produces almost exactly the same result
as UCCSD-VQE with the comparable ME of 19.87 kcal/mol. However, it should be kept in mind 
that ADAPT-SD and UCCSD are two different wave function ansatzes as discussed later.
%The total energy of UCCSD and ADAPT(3)-b 
%is minimized with respect to coefficients of one- and two-body excitation operators. As discussed 
%in traditional CC methods, the accuracy of UCCSD and ADAPT(3)-b can be systematically improved by
%including higher-order excitation operators.

UCCGSD-VQE gives a more promising ground state potential curve with the ME of only 1.58 kcal/mol. This
agrees with the conclusion revealed in the previous study 
that the UCCGSD ansatz is far more robust and accurate than the simple UCCSD ansatz~\cite{LeeHugHea19}. 
Analogous to ADAPT(3)-SD and UCCSD-VQE,
ADAPT(3) and UCCGSD-VQE also share the same operator pool and give quite close results. 
However, ADAPT-VQE requires much fewer parameters to achieve the accuracy of UCCGSD-VQE. 
This is mainly because ADAPT-VQE approximates 
the exact wave function with a compact sequence of unitary transformation acting on the reference state.
For example, the number of parameters in UCCGSD-VQE is fixed to be 236 while
ADAPT(3) requires at most 144 parameters with $R$(H-H) varying from 0.5 to 
2.0 \r{A}. This is achieved at the cost of computing {\it pre-estimated} gradients in Eq.~\eqref{eq:acse_res}. 

In addition, for ADAPT(3), the absolute energy error of the 1D hydrogen chain is 2-3 magnitude 
larger than those of small molecules (LiH, BeH$_2$ and H$_6$) presented in Ref.~\onlinecite{GriEcoBar19}.
As discussed in Section~\ref{sec:periodic}, these large deviations of the energy results from 
the complex wave function and Hamiltonian used in the VQE algorithm.
In order to analyze errors in detail,
Table~\ref{table1} shows the maximum absolute residual error (MARE) of
ACSE-Re and ACSE-Im for HF, UCCSD-VQE, UCCGSD-VQE and ADAPT(3). MAREs of ACSE-Im 
in UCCGSD-VQE and ADAPT(3) are very close to those in HF and UCCSD-VQE while 
MEs of the energy in UCCGSD-VQE and ADAPT(3) are much smaller than those in HF and UCCSD-VQE. 
This reveals that a majority of the absolute energy error in HF and UCCSD-VQE
originates from larger residual errors of ACSE-Re. 
In UCCGSD-VQE and ADAPT(3), the variational algorithm efficiently minimizes the MARE of ACSE-Re
($<$0.1 kcal/mol) while it is not able to simultaneously minimize the MARE of ACSE-Im, which is left 
to be as large as $\sim$5.2 kcal/mol. Therefore, the error of UCCGSD-VQE and ADAPT(3) are largely attributed to 
the MARE of ACSE-Im. In order to improve the accuracy of the quantum variational algorithm for periodic systems, 
it is necessary to remove the residual error of ACSE-Im.

%We also place one hydrogen atom in an unit cell and sample $k$-point with $1\times1\times8$, which 
%is expected to reproduce the similar results in Figure~\ref{fig:vqe_hx2}. 
%However, deviations of the total energy for UCCSD, UCCGSD, ADAPT(3)-b 
%and ADAPT(3) for comparison shown in Figure~\ref{fig:vqe_hx2} is quite different. 
%In this case, MAREs of HCSE happen to be very small and therefore the
%total energy minimized by the VQE algorithms converges towards the exact FCI energy.
%The MEs of the total energy are only 13.28, 0.06, 13.28, and 0.06 kcal/mol
%for UCCSD, UCCGSD, ADAPT(3)-b and ADAPT(3), respectively. 
%Therefore, the reference wave function have a significant influence on accuracy of 
%existing variational algorithms for the complex Hamiltonian. 
%Considering the $k$-point sampling Hamiltonian 
%in the calculations of periodic systems is in principle complex, the following discussions will focus on the 
%model with two hydrogen atoms and $1\times1\times4$ $k$-point sampling in an unit cell.

\begin{table*}[!tb] \label{table:energy_error}
\centering \caption{The mean error (ME) of the ground state energy (in kcal/mol) for various variational approaches.}
%\begin{tabular}{|p{2cm}|p{2cm}|p{2cm}|p{2cm}|p{2cm}|p{2cm}|p{2cm}|p{2cm}|p{2cm}|}
\begin{tabular}{|c|c|c|c|c|c|}
\hline 
 & HF & UCCSD-VQE & UCCGSD-VQE & ADAPT(3)-SD & ADAPT(1)    \\
 \hline
ME & 32.73 &	20.00 &	2.53 &	19.87 &	3.94  \\
\hline
& ADAPT(2)&ADAPT(3)& $\Gamma$-UCCSD-VQE &   $\Gamma$-ADAPT(3)-SD & $\Gamma$-ADAPT(X)-SD \\
\hline
ME & 2.63 & 2.94 & 1.69 &	1.20 & 0.07   \\
\hline
 &$\Gamma$-ADAPT(3) & ADAPT(3)-SD/QSE-SD &	ADAPT(3)-SD/QSE&	ADAPT(1)/QSE&	ADAPT(2)/QSE\\	
 \hline
 ME  &	0.05 &3.75 &	1.30 &	0.00 &	0.00 \\
 \hline
% &ADAPT(3)/QSE &&&& \\
% \hline
% ME & 0.00 &&&& \\
%hline 
\end{tabular}
\end{table*}

\subsection{VQE-K2G approach}

\begin{figure}[!htb]
\begin{center}
\includegraphics[width=0.45\textwidth]{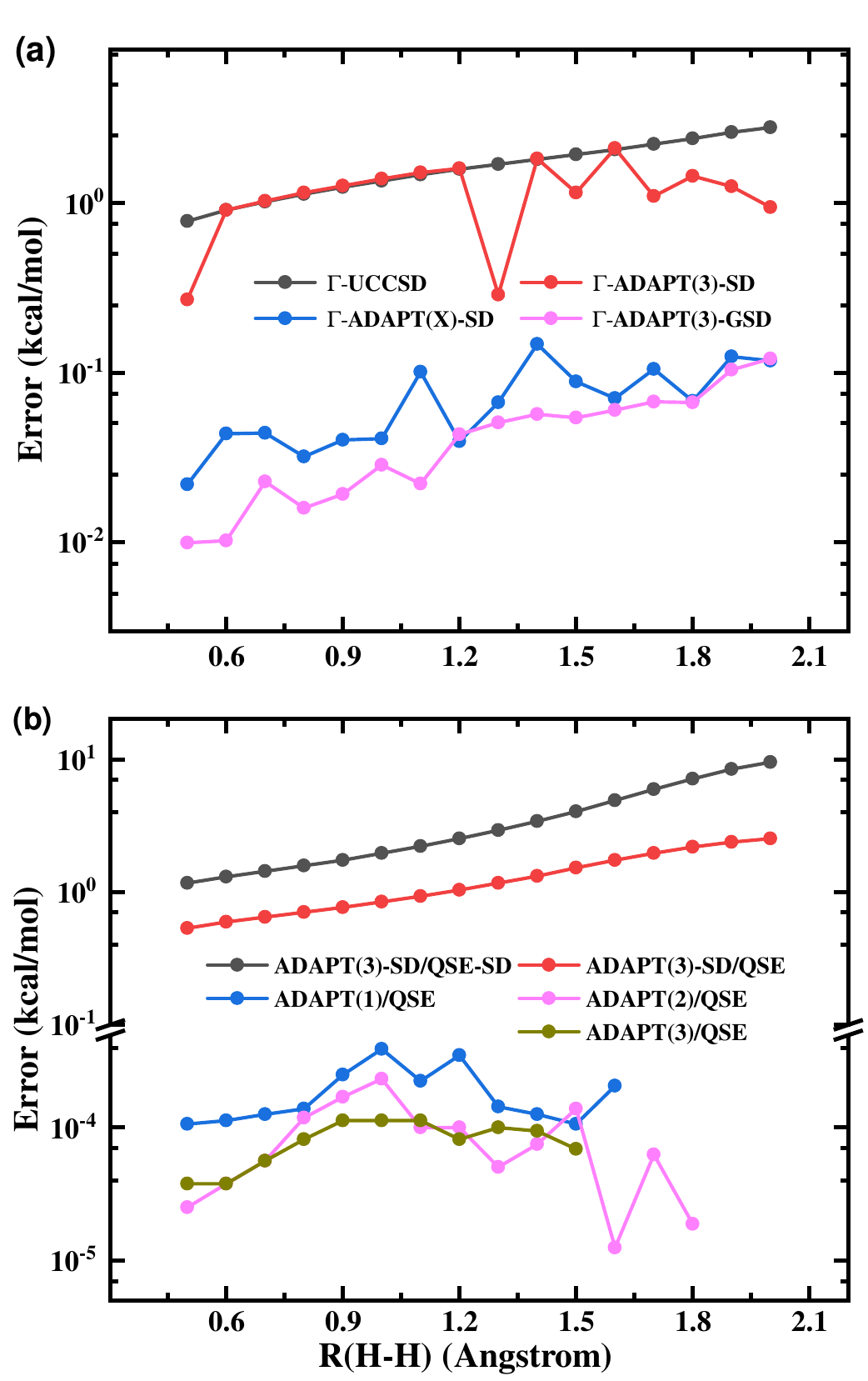}
\end{center}
\caption{Error in the absolute energy of the VQE-K2G approach (a) and the VQE/QSE 
approach (b) for the ground state of 1D hydrogen chain. $\Gamma$-ADAPT(X)-SD 
indicates $\epsilon$=$2\times10^{-4}$. Vanishing points 
indicate zero error when the energies are recorded with 
eight decimal places in Hartree.} \label{fig:k2gamma}
\end{figure}

%A straightforward generalization of the existing VQE algorithms to the periodic systems
%will lead to an uncontrollable residual error of ACSE-Im.
%Here, we assess the accuracy of the K2Gamma scheme, which systematically remove 
%the residual error of ACSE-Im through an unitary transformation of molecular orbitals.
In Figure~\ref{fig:k2gamma}(a), we show the absolute energy error of various VQE-K2G approaches 
as a function of $R$(H-H), which systematically remove 
the residual error of ACSE-Im through an unitary transformation of HF orbitals at sampling $k$-points.
%in which the molecular orbitals is expanded with the $\Gamma$-point BAOs of eight 
%hydrogen atoms in an unit cell. 
Analogous to molecular simulations, the potential energy curve computed with $\Gamma$-ADAPT(3) agrees
quite well with the exact curve since the variationally optimized wave function strictly satisfies 
ACSE with the norm of residual errors less than $1\times10^{-3}$.
ADAPT-VQE is an adaptive algorithm and the wave function ansatz is self-consistently grown. 
Therefore, when the potential energy curve is scanned, the energy discontinuity may appear 
if the convergence thresh is not small enough. 
%In addition, the operator pool in $\Gamma$-ADAPT(m)-SD
%is a subset of that in $\Gamma$-ADAPT(m). This implies that $\Gamma$-ADAPT(m)-SD converges with 
%a sub-criteria of $\Gamma$-ADAPT(m). As a result, $\Gamma$-ADAPT(m)-SD will need a much tighter convergence thresh 
For example, errors in the absolute energy computed with $\Gamma$-ADAPT(3)-SD dramatically 
fluctuates as shown in Figure~\ref{fig:k2gamma}(a). 
Here, we also compute the energy with a tighter convergence thresh 
$\epsilon$=$2\times10^{-4}$ in $\Gamma$-ADAPT(m)-SD, named $\Gamma$-ADAPT(X)-SD, for comparison. 
$\Gamma$-ADAPT(X)-SD gives well converged results with 
ME of 0.07 kcal/mol, which is even comparable to ME of 0.05 kcal/mol in $\Gamma$-ADAPT(3).
%As discussed in Ref.~\onlinecite{GriEcoBar19}, 
%The variationally optimized wave function in $\Gamma$-ADAPT(m) 
%satisfies ACSE up to the convergence thresh $\epsilon$, which means that the 
%exact wave function can be well approximated by the $\Gamma$-ADAPT(m) wave function 
%with only one- and two-body general excitation operators.

In comparison with UCCSD-VQE, $\Gamma$-UCCSD-VQE gives a much better description of the potential energy 
curve of 1D hydrogen chain. The maximum deviation of the energy for $\Gamma$-UCCSD-VQE at $R=2.0$ \r{A}
is only 2.75 kcal/mol. Different from what is shown in Figure~\ref{fig:vqe_hx1}, 
$\Gamma$-ADAPT(3)-SD and $\Gamma$-ADAPT(X)-SD give more accurate results than $\Gamma$-UCCSD-VQE. 
In the ADAPT-VQE method, high-order excitations are ultimately included
after a sequence of low-order exponentiated excitation operators acting on the wave function. 
A tighter convergence thresh implies that a larger $N_k$ in Eq.~\ref{eq:wave0} is required to converge the wave function ansatz. 
As $N_k$ increases, the ADAPT-VQE wave function is in principle
expanded in a larger configuration state space, which will improve the accuracy of the wave function ansatz.
Therefore, the absolute energy error of $\Gamma$-ADAPT(m)-SD is reduced as $m$ increases. This is very similar with the $k$-UpCCGSD
approach, in which a larger $k$ was often used to improve the accuracy~\cite{LeeHugHea19}. 
UCCSD truncates excitation operators up to the second order and its wave function is 
optimized in a fixed parameter space. The inclusion of higher-order excitation operators is 
the most straightforward way to improve the accuracy of $\Gamma$-UCCSD while  
at a steeply increasing computational cost. 

It is worthy to mention that the accuracy of ADAPT(m)-SD can not be improved by increasing $m$.
The main reason is the limitation of crystal momenta conservation in excitation operators
when HF orbitals at sampling k-points are used. Given excitation operators in Eq.~\eqref{eq:pbc_t},
it is impossible to decompose these high-order excitations into the product of low-order excitation operators 
with crystal momentum conservation. For example, a quadruple excitation operator, 
\begin{equation}
    \hat{T}^{\tilde{p}_1\tilde{p}_2\tilde{p}_3\tilde{p}_4}_{\tilde{q}_1\tilde{q}_2\tilde{q}_3\tilde{q}_4} = a_{\tilde{p}_1}^\dag a_{\tilde{p}_2}^\dag a_{\tilde{p}_3}^\dag a_{\tilde{p}_4}^\dag a_{\tilde{q}_1}
    a_{\tilde{q}_2}a_{\tilde{q}_3}a_{\tilde{q}_4}
\end{equation}
conserves crystal momentum
\begin{equation}\label{eq:quad}
    \sum_{i=1}^4 \bk_{\tilde{p}_i} - \sum_{j=1}^4 \bk_{\tilde{q}_j} = \mathbf{G}_m.
\end{equation}
Eq.~\ref{eq:double} is only sufficient but not necessary condition of Eq.~\eqref{eq:quad}.
Therefore, ADAPT(m)-SD optimizes the wave function in a limited subspace of configuration states,
which can not be simply overcome by increasing $m$. Analogous to UCCSD, the explicit 
inclusion of higher-order excitations is necessary to improve the 
accuracy of ADAPT-SD.

\begin{figure}[!htb]
\begin{center}
\includegraphics[width=0.45\textwidth]{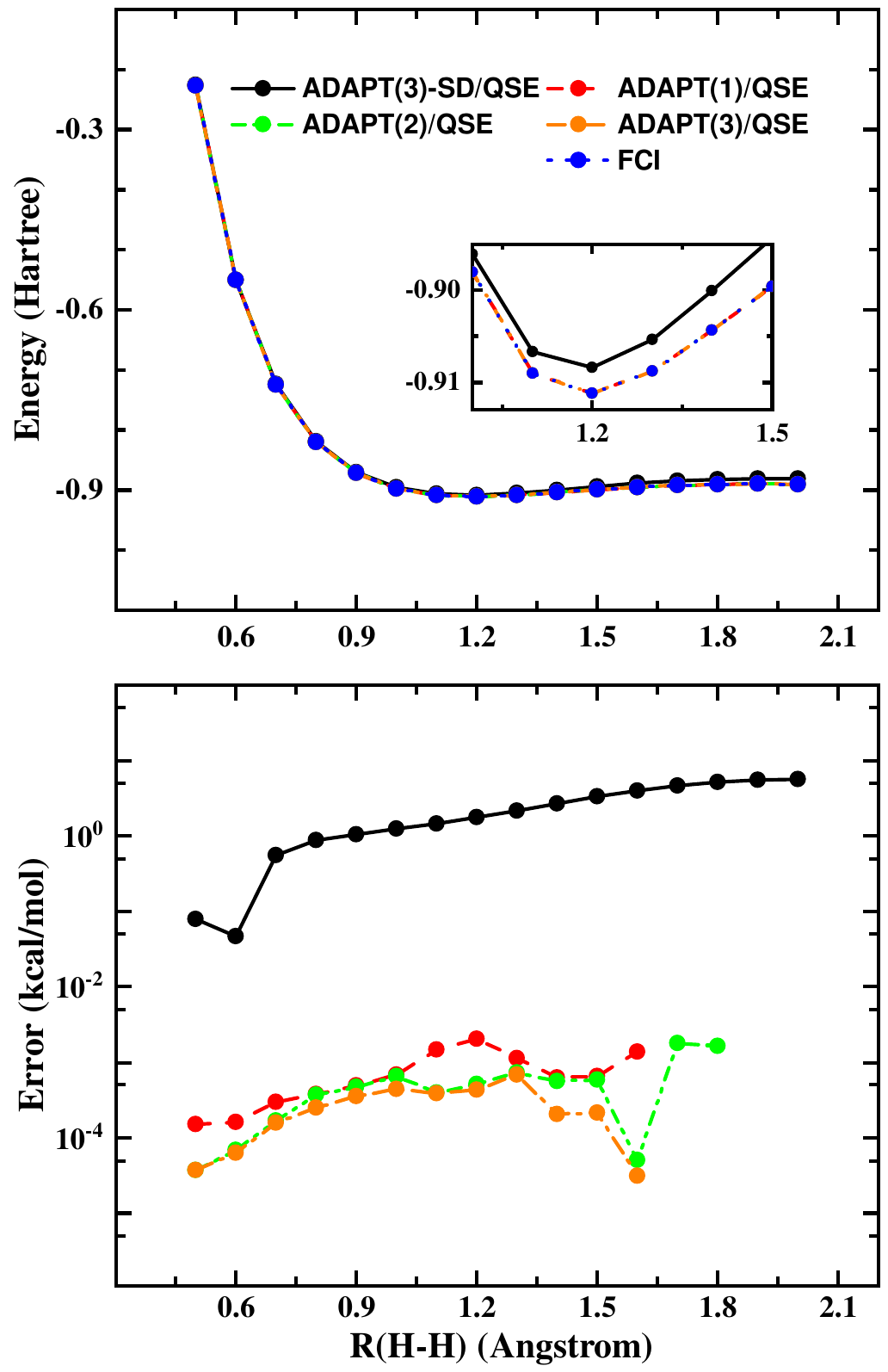}
\end{center}
\caption{The first excited-state potential energy curve and the absolute energy error of 
1D hydrogen chain computed with ADAPT-VQE/QSE. Vanishing points 
indicate zero error when the energies are recorded with 
eight decimal places in Hartree.} \label{fig:qse}
\end{figure}

\subsection{Quantum subspace expansion}

The quantum expansion subspace method is an alternative approach to improve the estimation of
the energy over the reference state. Here, VQE/QSE can be considered as a {\it Post}-VQE method to improve the VQE algorithm. 
In Figure~\ref{fig:k2gamma}(b), we present the absolute energy error of the VQE/QSE approach
as a function of $R$(H-H).
Here, the reference states are prepared with ADAPT(3)-SD, ADAPT(1), ADAPT(2) 
and ADAPT(3). The Hamiltonian matrix elements are sampled in the linear-response space of 
the VQE wave function with single and double excitations. 
All energy curves computed with ADAPT(m)/QSE exactly reproduce the FCI result
with errors less than $1\times10^{-3}$ kcal/mol. This demonstrates that ADAPT(m) is able to obtain 
a quite reasonable reference state even with a very large convergence thresh, such as $\epsilon$=$1\times10^{-1}$.
%which indicates few parameters to prepare the wave function. 
Although the mean error of 1.30 kcal/mol in ADAPT(3)-SD/QSE
gets slightly beyond the chemical accuracy, this is almost 3-4 magnitude larger than MEs of ADAPT(m)/QSE. 
Since the Hamiltonian matrix elements are sampled in the 
same linear-response space, the reference state prepared with ADAPT(3)-SD is even worst than ADAPT(1). While
the number of parameters in both ADAPT(3)-SD and ADAPT(1) is 30.
For QSE-SD, the Hamiltonian is sampled in a much
smaller linear response space of dimension being 153 (625 for QSE). This leads to a remarkable deviation of 
the potential energy curve for ADAPT(3)-SD/QSE-SD with the ME of 3.75 kcal/mol. Therefore, the accuracy of the QSE approach 
strongly depend on the reference state and the truncation of excitation operators. 
Sometimes higher-order excitations should be included in order to obtain a converged result.

As the substantial success of the VQE algorithm in simulating ground-state properties, there is a broad
interest in applying it to study excited states. For the 1D hydrogen chain, 
the potential energy surface of different electronic states is important to understand 
the rich and fascinating phase diagram in quantum material physics.
In Figure \ref{fig:qse}, we present the first excited state energy and the absolute energy error of 1D hydrogen chain
as a function of $R$(H-H).
The ground state wave function is obtained with ADAPT(m) or ADAPT(m)-SD and the first excited state is computed with the QSE approach.
Regardless of the convergence thresh $\epsilon$ used in ground-state ADAPT(m) calculations, we obtain
exactly the same curves as FCI. For ADAPT(m), mean errors of the first excited state energy are 
$5.91\times 10^{-4}$, $5.05\times 10^{-4}$, $2.04 \times 10^{-4}$ kcal/mol for
$\epsilon$=$10^{-1}$, $\epsilon$=$10^{-2}$ and $\epsilon$=$10^{-3}$, respectively. The performance of 
ADAPT(3)-SD/QSE in excited-state calculations is quite similar with that in the ground state calculation. 
Although the absolute energy error of ADAPT(3)-SD/QSE is acceptable, we recommend to prepare the ground state with ADAPT(m)
since it has been validated to be more robust and accurate.

\section{Conclusion}\label{sec:conclusion}

In this work, we generalize the variational quantum eigenvalue algorithm for periodic systems. 
We first carry out classical VQE simulations of 1D infinite hydrogen chain
with UCC and ADAPT ansatzes using HF orbitals at sampling $k$-points.
UCCSD-VQE and ADAPT(3)-SD totally fails to accurately describe the potential energy surface of 1D hydrogen chain. 
The absolute energy error of UCCGSD and ADAPT(3) is acceptable while it is at least 1-2 magnitude 
larger than that in the molecular simulation. The detailed analysis of residual error of 
ACSE reveals that the significant deviation of the energy results from the complex 
wave function and Hamiltonian generated based on HF orbitals at sampling $k$-points. 

Then, we present two schemes to overcome this problem in the VQE algorithm for periodic systems. 
One is the VQE-K2G approach, which avoids the complex wave function and Hamiltonian involved in VQE 
through an unitary transforming of HF orbitals at sampling $k$-points in an unit cell into real 
orbitals at $\Gamma$-point in a supercell. The VQE-K2G algorithm totally removes 
the residual error of imaginary part of ACSE and then achieve the same accuracy as the VQE algorithm in molecular simulations.  
Another scheme is the combination of VQE with the quantum subspace expansion approach, which offers a better
estimation of the energy over the reference state. The VQE/QSE approach projects the Hamiltonian onto the linear
response space of the reference state prepared with the VQE algorithm. This can be considered 
as a {\it Post}-VQE correction to the reference state energy. Numerical simulations demonstrates that
both the VQE-K2G and VQE/QSE approaches provides an accurate enough 
description of the potential energy surface of 1D hydrogen chain. In addition, the accuracy of 
the VQE/QSE approach is 1-2 magnitude smaller than that of the VQE-K2G approach at the expense of 
steeply increasing measurements. 

In this work, we come to the same conclusion that
UCCGSD is more stable than UCCSD as stated in previous studies.
The accuracy of UCCGSD without Trotterization is comparable to ADAPT. 
As Trotterization is introduced in UCC, the wave function expression of Eq.~\eqref{eq:ucc_trotter} is 
quite similar with that in ADAPT as shown in Eq.~\eqref{eq:wave0}. However, UCCGSD uses the same 
coefficients of excitation operators in each Trotterization step while ADAPT uses difference 
coefficients. Therefore, ADAPT is expected to be more flexible than UCCGSD with Trotterization.
In addition, ADAPT generates a optimized sequence of unitary transformation in Eq.~\eqref{eq:wave0},
which is proven to be necessary to find the lowest energy~\cite{GriClaEco20}. 
However, at the beginning of each iteration, ADAPT-VQE needs to compute the gradients in Eq.~\eqref{eq:acse_grad}
at the expense of $N_b^8$ measurements where $N_b$ is the number of basis functions. 
Further work should be devoted to reduce the number of 
measurements.

Finally, we note that the wave function ansatz based on Gaussian basis functions offer a
well established solution of electronic structure problems. While a large Gaussian basis set
is often required to obtain the converged result. In order to 
accurately simulate electron structure properties in material science, an alternative approach is to use 
the Wannier basis function~\cite{MarMosYat12} or adaptive local basis function in a discontinuous 
Galerkin framework~\cite{ZhaLinHu17}, which can reach the complete basis set limit with much fewer 
basis functions.

\section{Acknowledgments}
This work is supported by the National Natural Science
Foundation of China (21688102,21825302,21803065),
by National Key Research and Development
Program of China (2016YFA0200604), Anhui Initiative in
Quantum Information Technologies (AHY090400).
%\bibliographystyle{acs}
%\bibliography{qc}

\providecommand{\refin}[1]{\\ \textbf{Referenced in:} #1}

\end{document}